\documentclass[11pt,a4paper]{article}

\usepackage{amsmath,amssymb}
\usepackage{bm}
\usepackage{mathrsfs}
\usepackage{multirow}
\usepackage{float}
\usepackage[usenames]{color}
\usepackage{indentfirst}
\usepackage{cite}
\usepackage{cases}
\usepackage{geometry}
\usepackage{comment}
\usepackage[dvipdfmx]{graphicx}
\usepackage{simplewick}
\usepackage{mathtools}

\newcommand{\fsl}[1]{\ensuremath{\mathrlap{\!\not{\phantom{#1}}}#1}}

\allowdisplaybreaks[4]
\makeatletter
\@addtoreset{equation}{section}

\makeatother

\begin{document}

\title{Graviton loop contribution to Higgs potential \\in gauge-Higgs unification}
\author{
Yasunari \textsc{Nishikawa}\footnote{E-mail: 17st308a@shinshu-u.ac.jp}\\
{\it Department of Physics, Shinshu University, }\\
{\it Matsumoto 390-8621, Japan}
}
\date{}
\maketitle

\begin{abstract}
We study a two-loop finiteness of an effective potential for a Higgs boson
that is the fifth component of a gauge field in an $U(1)$ gauge theory coupled to quantum gravity
on the five-dimensional space-time $M^4\times S^1$.
There are two types of diagrams including quantum gravitational corrections.
We find that only one type of diagram contributes to the effective potential for the Higgs boson in fact
and its magnitude is finite.
\end{abstract}

\section{Introduction}
\label{sec:Intro}
Scalar fields perform very crucial roles in particle physics and cosmology.
In the standard model (SM), the Higgs boson gives fermions and gauge bosons their masses through the Higgs mechanism.
In slow-roll inflation models, an exponential expansion of space is realized by a suitable potential of an inflaton.
However, a scalar field causes a serious problem that its mass parameter is very sensitive to a cutoff scale.
There is no principle to protect it from huge quantum corrections, so that we need unnatural fine-tuning in both models.
Several solutions have been proposed using supersymmetry, composite Higgs models, and higher-dimensional theories.
Although any indications of such new physics have not yet been found at Large Hadron Collider (LHC), they are still attractive.
In this paper, we focus on higher-dimensional theories.

Higher-dimensional theories were originally proposed to unify four-dimensional (4D) Einstein gravity and electromagnetism
into five-dimensional (5D) Einstein gravity\cite{Kaluza:1921tu}\cite{Klein:1926tv}.
Although the original attempts have failed, the basic idea to unify 4D fields into a higher-dimensional field
survives as a useful concept even today.
Particularly, gauge-Higgs unification (GHU) models have been well investigated
\cite{Manton:1979kb}\cite{Fairlie:1979at}\cite{Fairlie:1979zy}.
In these models, Wilson line phases are defined using zero modes of extra-dimensional components
of a higher-dimensional gauge field, and regarded as scalar fields.
They are massless at the tree level due to the gauge invariance,
namely, their masses are expected to be free from the ultra-violet (UV) divergence.
On the other hand, finite effective potentials for the scalar fields are predicted to be generated by quantum corrections.
It comes from a conjecture that the potentials can be induced as functions for the Wilson loops,
which respect the gauge symmetry, perturbatively.
For these reasons, the scalar fields can gain finite masses without fine-tuning.
The effective potentials also trigger a spontaneous breakdown of symmetries,
which is called the Hosotani mechanism\cite{Hosotani:1983xw}\cite{Hosotani:1983vn}\cite{Hosotani:1988bm}.
If the scalar field is identified as the Higgs boson of the SM, the gauge hierarchy problem can be solved
and the electroweak symmetry can be broken through the Hosotani mechanism\cite{Hatanaka:1998yp}.
Similarly, it is regarded as the inflaton in extranatural inflation models,
so that the fine-tuning problem for the slow-roll potential can be solved\cite{ArkaniHamed:2003wu}.

In a sense, \textit{the raison d'\^etre} of these models depends on a conjecture
that the effective potentials for the scalar fields are finite in spite of a non-renormalizability of higher-dimensional theories.
There are several arguments on the finiteness,
but it has not yet been possible to confirm it\cite{Hosotani:2005fk}\cite{Hosotani:2006nq}.
In both Abelian and non-Abelian gauge theories without graviton, it is verified at the two-loop level at present
\cite{Shiraishi:1991hs}\cite{Maru:2006wa}\cite{Hosotani:2007kn}\cite{Hisano:2019cxm}.
Although graviton hardly affects phenomenological analyses because of tininess of the coupling constant,
it is meaningful to study whether the effective potentials for the scalar fields are finite or not
in higher-dimensional gauge theories with graviton.
If the finiteness is broken by graviton,
the advantage of the GHU models is spoiled just by coupling quantum gravity (QG) to the models.
Then, the GHU models might have a serious weakness.
Since a graviton has no internal gauge quantum numbers, it contributes to the potentials for the first time at the two-loop level.
Thus, we need to verify the finiteness at the two-loop level in the cases with graviton.

In this paper, we investigate a two-loop finiteness of an effective potential for a Higgs boson
in an $U(1)$ gauge theory coupled to QG on the 5D space-time $M^4\times S^1$.
Here and hereafter, we refer to the scalar field that originates from a 5D $U(1)$ gauge boson as Higgs boson.

The content of our paper is as follows.
In the next section, we review the effective potential for the Higgs boson up to the two-loop level
in an $U(1)$ gauge theory on $M^4\times S^1$.
In Sect.\,3, we evaluate graviton contributions to the two-loop effective potential.
In the last section, we give conclusions and discussions.

\section{$U(1)$ gauge theory on $M^4\times S^1$}
\label{sec:U(1)}
In this section, we review the effective potential for the Higgs boson up to the two-loop level
in an $U(1)$ gauge theory defined on $M^4\times S^1$\cite{Hosotani:2007kn}.
Here, $M^4$ is a 4D Minkowski space and $S^1$ is a circle with a radius $R$,
whose coordinates are denoted by $x^\mu$ ($\mu=0, 1, 2, 3$) (or $x$) and $y$ ($=x^5$), respectively.
We also use the 5D notation $x^M$ ($M=0, 1, 2, 3, 5$) (or $\hat{x}$).

The 5D Lagrangian density is given by
\begin{align}
\label{5dlpe}
\hat{\mathcal{L}}_{\mathrm{eff}}
=-\frac{1}{4}F_{MN}F^{MN}+\bar{\psi}i\Gamma^MD_M\psi-\frac{1}{2}(\partial_MA^M)^2.
\end{align}
Here, the covariant derivative acting on $\psi$ is $D_M=\partial_M+ieA_M$,
and the 5D gamma matrices $\Gamma^M$ are given by
\begin{align}
\label{5gm}
\Gamma^{\mu}=\gamma^{\mu},~ \Gamma^5=i\gamma^5,
\end{align}
using the 4D gamma matrices $\gamma^\mu$ and $\gamma^5=i\gamma^0\gamma^1\gamma^2\gamma^3$.
They satisfy the algebraic relation $\{\Gamma^M,\Gamma^N\}=2\eta^{MN}$ with $\eta^{MN}=\mathrm{diag}(1, -1, -1, -1, -1)$.
From the requirement that physical quantities should be single-valued functions of space-time coordinates,
the boundary conditions of the fields on $S^1$ are determined as,
\begin{align}
\label{bc1}
A_M(x,y+2\pi R)=A_M(x,y),~~
\psi(x,y+2\pi R)=e^{i\alpha}\psi(x,y),
\end{align}
where $\alpha$ is an arbitrary constant phase.
For simplicity, we fix the phase as $\alpha=0$ in this paper.
Then, the fields are given by Fourier expansions and only zero modes are observed as 4D fields.
Especially, we obtain the Higgs boson defined as the Wilson line phase $\theta$ from the fifth component of $A_M$ such that
\begin{align}
\label{wlp}
\theta=e\int_{-\pi R}^{\pi R}dy\frac{1}{\sqrt{2\pi R}}A_5^{(0)}=2\pi Re_4A_5^{(0)},
\end{align}
where $A_5^{(0)}$ denotes the zeroth Fourier coefficient of $A_5(x,y)$,
and $e_4(=e/\sqrt{2\pi R})$ is the 4D gauge coupling constant.

To compute the effective potential for $\theta$, we separate $A_5$ into the classical and quantum fields as
$A_5\rightarrow A_5^\mathrm{c}+A_5$.
Hence, the Lagrangian density (\ref{5dlpe}) is rewritten as
\begin{align}
\label{5dlpe2}
\hat{\mathcal{L}}_{\mathrm{eff}}
=\frac{1}{2}A_M\eta^{MN}\hat{\Box}A_N+\bar{\psi}i\Gamma^M\tilde{\partial}_M\psi
-e\bar{\psi}\Gamma^MA_M\psi,
\end{align}
where $\tilde{\partial}_M=\partial_M+ie\delta_M^5A_5^\mathrm{c}$.
Then, a five-momentum for $\psi$ is given by $\tilde{p}_M=(p_\mu,p_5-\theta/2\pi R)=(p_\mu,m/R-\theta/2\pi R)$
where $m$ is an integer.
The fifth component of the momentum is discretized because of the compact space $S^1$,
and we pay attention to only the zero mode of $A_5^\mathrm{c}$.
On the other hand, a five-momentum for $A_M$ is given by $k_M=(k_\mu,k_5)=(k_\mu,\ell/R)$
where $\ell$ is an integer.
As a result, only fermion contributes to the one-loop effective potential:
\begin{align}
\label{5d1p}
-i\hat{V}^{1\mathchar`-\mathrm{loop}}
&=2i\int \frac{d^5p_{\mathrm{E}}}{(2\pi)^5}\ln (\hat{\tilde{p}}_{\mathrm{E}}^2) \nonumber \\
&=\frac{2i}{2\pi R}\sum_{m=-\infty}^{\infty}\int \frac{d^4p_{\mathrm{E}}}{(2\pi)^4}
\ln\biggl\{p_{\mathrm{E}}^2+\biggl(\frac{m}{R}-\frac{\theta}{2\pi R}\biggr)^2\biggr\},
\end{align}
where subscript $\mathrm{E}$ stands for Euclidean quantities.
In the second equality, we change the fifth momentum integral into the infinite summation.
The effective potential is expected to be a finite periodic function for $\theta$.
In fact, we can separate $\theta$-independent divergences from the expression (\ref{5d1p})
using the Poisson resummation formula, and hence the 4D one-loop effective potential is given by
\begin{align}
\label{4d1p}
V^{1\mathchar`-\mathrm{loop}}(\theta)
=\frac{3}{16\pi^6R^4}\sum_{m=1}^{\infty}\frac{\cos(m\theta)}{m^5},
\end{align}
where $V^{1\mathchar`-\mathrm{loop}}=2\pi R\hat{V}^{1\mathchar`-\mathrm{loop}}$.
Obviously, $V^{1\mathchar`-\mathrm{loop}}$ has a minimum at $\theta=\pi$.

\begin{figure}[t]
\begin{center}
\includegraphics[width=2cm]{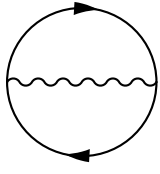}
\end{center}
\caption{1-PI vacuum diagram including $U(1)$ gauge boson and fermions}
\label{fig:pe}
\end{figure}

Next, we compute the two-loop effective potential.
It is represented by the sum of one-particle irreducible (1-PI) vacuum diagrams involving fermions.
In our case, it is given by the diagram in Figure\,\ref{fig:pe} where wiggly and solid lines denote $A_M$ and $\psi$, respectively.
It is expressed as
\begin{align}
\label{5d2ppe}
&-i\hat{V}_{\mathrm{Fig}.\,\ref{fig:pe}}^{2\mathchar`-\mathrm{loop}} \nonumber \\
&=-\frac{(-ie)^2}{2!}\int \frac{d^5k}{(2\pi)^5}\int \frac{d^5p}{(2\pi)^5}\int \frac{d^5q}{(2\pi)^5}
(2\pi)^5\delta^{(5)}(\hat{k}+\hat{p}-\hat{q})
\tilde{D}_{MM'}(\hat{k})
\mathrm{tr}\bigl\{\Gamma^M\tilde{S}(\hat{\tilde{p}})
\Gamma^{M'}\tilde{S}(\hat{\tilde{q}})\bigr\} \nonumber \\
&=6ie^2\int \frac{d^5k_{\mathrm{E}}}{(2\pi)^5}
\int \frac{d^5p_{\mathrm{E}}}{(2\pi)^5}\frac{1}{\hat{k}_{\mathrm{E}}^2\hat{\tilde{p}}_{\mathrm{E}}^2}
-3ie^2\int \frac{d^5p_{\mathrm{E}}}{(2\pi)^5}
\int \frac{d^5q_{\mathrm{E}}}{(2\pi)^5}\frac{1}{\hat{\tilde{p}}_{\mathrm{E}}^2\hat{\tilde{q}}_{\mathrm{E}}^2} \nonumber \\
&=\frac{6ie^2}{(2\pi R)^2}\sum_{\ell=-\infty}^{\infty}\sum_{m=-\infty}^{\infty}
\int \frac{d^4k_{\mathrm{E}}}{(2\pi)^4}\int \frac{d^4p_{\mathrm{E}}}{(2\pi)^4}
\frac{1}{\{k_{\mathrm{E}}^2+(\frac{\ell}{R})^2\}\{p_{\mathrm{E}}^2+(\frac{m}{R}-\frac{\theta}{2\pi R})^2\}} \nonumber \\
&~~~~~~
-\frac{3ie^2}{(2\pi R)^2}\sum_{m=-\infty}^{\infty}\sum_{n=-\infty}^{\infty}
\int \frac{d^4p_{\mathrm{E}}}{(2\pi)^4}\int \frac{d^4q_{\mathrm{E}}}{(2\pi)^4}
\frac{1}{\{p_{\mathrm{E}}^2+(\frac{m}{R}-\frac{\theta}{2\pi R})^2\}\{q_{\mathrm{E}}^2+(\frac{n}{R}-\frac{\theta}{2\pi R})^2\}},
\end{align}
where $\tilde{D}_{MM'}(\hat{k})$ and $\tilde{S}(\hat{\tilde{p}})$ are propagators defined
in Appendix \ref{sec:ap1}.
We can rewrite the infinite sums using the formula (\ref{tif1}), and hence the 4D two-loop effective potential is given by
\begin{align}
\label{4d2ppe}
V_{\mathrm{Fig}.\,\ref{fig:pe}}^{2\mathchar`-\mathrm{loop}}(\theta)
=-\frac{6e_4^2}{16\pi^4(2\pi R)^4}\sum_{\ell,m=1}^{\infty}\frac{1}{\ell^3}\frac{\cos(m\theta)}{m^3}
+\frac{3e_4^2}{16\pi^4(2\pi R)^4}\sum_{m,n=1}^{\infty}\frac{\cos(m\theta)}{m^3}\frac{\cos(n\theta)}{n^3},
\end{align}
where $V_{\mathrm{Fig}.\,\ref{fig:pe}}^{2\mathchar`-\mathrm{loop}}
=2\pi R\hat{V}_{\mathrm{Fig}.\,\ref{fig:pe}}^{2\mathchar`-\mathrm{loop}}$
and terms with $\theta$-independent divergences are dropped.
$V_{\mathrm{Fig}.\,\ref{fig:pe}}^{2\mathchar`-\mathrm{loop}}$ is
the finite periodic function for $\theta$ as well as $V^{1\mathchar`-\mathrm{loop}}$.

\section{$U(1)$ gauge theory coupled to QG on $M^4\times S^1$}
\label{sec:U(1)QG}
In this section, we evaluate the contributions to the effective potential from QG on the background $M^4\times S^1$.
Since a graviton has no $U(1)$ charge, it does not contribute to the potential at the one-loop level.
Thus, we calculate the two-loop contributions including gravitons and fermions as intermediate states.

We consider a 5D $U(1)$ gauge theory coupled to an Einstein gravity.
The Lagrangian density of the theory is given by
\begin{align}
\label{5dlge}
\hat{\mathcal{L}}_{\mathrm{eff}}=\sqrt{g}\biggl(-\frac{1}{\kappa^2}\mathcal{R}
+\frac{1}{2}\bar{\psi}ie^M_{\phantom{M}a}\Gamma^aD_M\psi
-\frac{1}{2}\bar{\psi}\overleftarrow{D}_Mie^M_{\phantom{M}a}\Gamma^a\psi\biggr)+\hat{\mathcal{L}}_{\mathrm{GF}},
\end{align}
where indices written by uppercase and lowercase letters denote covariant and local Lorentz indices,
a flat metric is $\mathrm{diag}(1, -1, -1 ,-1, -1)$,
$g=\mathrm{det}(g_{MN})$, $\kappa$ is the 5D gravitational coupling constant,
$\mathcal{R}$ is a Ricci scalar,  and $e^M_{\phantom{M}a}$ is a f\"{u}nfbein, respectively.
$\hat{\mathcal{L}}_{\mathrm{GF}}$ is a gauge fixing term.
The covariant derivative acting on $\psi$ is given by
\begin{align}
\label{cdae}
D_M\psi=\biggl(\tilde{\partial}_M-\frac{1}{4}\omega_M^{bc}\Gamma_{bc}\biggr)\psi,~~
\bar{\psi}\overleftarrow{D}_M=
\tilde{\partial}_M^*\bar{\psi}+\bar{\psi}\biggl(\frac{1}{4}\omega_M^{bc}\Gamma_{bc}\biggr),
\end{align}
with
\begin{align}
\omega_{M}^{bc}=
-g^{NS}e_{S}^{\phantom{S}b}(\partial_Me_{N}^{\phantom{N}c}-\Gamma_{\phantom{P}MN}^Pe_{P}^{\phantom{P}c}),~~
\Gamma_{bc}=\frac{1}{2}[\Gamma_b,\Gamma_c],
\end{align}
where $\Gamma_{\phantom{P}MN}^P=g^{PQ}(\partial_Ng_{QM}+\partial_Mg_{QN}-\partial_Qg_{MN})/2$,
and $*$ stands for the complex conjugation.
We ignore terms involving $A_M$ in $\hat{\mathcal{L}}_{\mathrm{eff}}$, $D_M$, and $\overleftarrow{D}_M$,
since they do not concern our purpose.
The f\"{u}nfbein satisfies the periodic boundary condition on $S^1$.

To perform perturbative calculations, we separate the metric into the Minkowski background and the quantum fluctuation
as $g_{MN}=\eta_{MN}+\kappa h_{MN}$.
Then, $g^{MN}$ and $\sqrt{g}$ are expanded up to $\mathcal{O}(h^2)$ as
\begin{align}
\label{me}
g^{MN}=\eta^{MN}-\kappa h^{MN}+\kappa^2h^{MP}h_{\phantom{N}P}^N,~~
\sqrt{g}=1+\frac{1}{2}\kappa h-\frac{1}{4}\kappa^2h_{MN}h^{MN}+\frac{1}{8}\kappa^2h^2,
\end{align}
where $h=h_M^{\phantom{M}M}$.
Similarly, $e_M^{\phantom{M}a}$ and $e_{\phantom{M}a}^M$ are done \cite{Donoghue:2015hwa}:
\begin{align}
\label{ve}
e_M^{\phantom{M}a}=\delta_M^a+\frac{1}{2}\kappa h_{\phantom{a}M}^a-\frac{1}{8}\kappa^2h_{MP}h^{aP},~~
e_{\phantom{M}a}^M=\delta_a^M-\frac{1}{2}\kappa h_{\phantom{M}a}^M+\frac{3}{8}\kappa^2h_{aP}h^{MP}.
\end{align}
After these expansions, it is not necessary to distinguish between covariant and local Lorentz indices.
In the following, we use only indices written by uppercase letters.

\begin{figure}[t]
\begin{minipage}{0.5\hsize}
\begin{center}
\includegraphics[width=2cm]{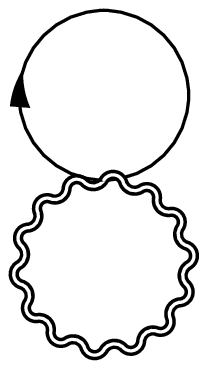}
\end{center}
\end{minipage}
\begin{minipage}{0.5\hsize}
\begin{center}
\includegraphics[width=2cm]{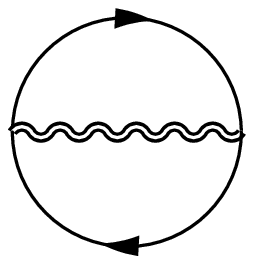}
\end{center}
\end{minipage}
\caption{Two-loop vacuum diagrams including graviton and fermions}
\label{fig:ge}
\end{figure}

We expand the Lagrangian density (\ref{5dlge}) up to $\mathcal{O}(h^2)$,
using the relations $g_{MN}=\eta_{MN}+\kappa h_{MN}$, (\ref{me}), and (\ref{ve}), as
\begin{align}
\label{5dlge2}
\hat{\mathcal{L}}_{\mathrm{eff}}
=-\frac{1}{2}h_{MN}\biggl\{\frac{1}{4}\biggl(\eta^{MP}\eta^{NQ}+\eta^{MQ}\eta^{NP}
-\eta^{MN}\eta^{PQ}\biggr)\biggr\}\hat{\Box}h_{PQ}
+\bar{\psi}i\Gamma^M\tilde{\partial}_M\psi
+\hat{\mathcal{L}}_{\psi}^{(1)}+\hat{\mathcal{L}}_{\psi}^{(2)},
\end{align}
where $\hat{\mathcal{L}}_{\psi}^{(1)}$ and $\hat{\mathcal{L}}_{\psi}^{(2)}$ are interaction terms given
in Appendix \ref{sec:ap3}, and the gauge fixing term is given by
\begin{align}
\label{gft}
\hat{\mathcal{L}}_{\mathrm{GF}}=\frac{1}{2}\biggl(\partial^Nh_{MN}-\frac{1}{2}\partial_Mh\biggr)
\biggl(\partial_Ph^{MP}-\frac{1}{2}\partial^Mh\biggr).
\end{align}
Since three and four-points interactions are found in $\hat{\mathcal{L}}_{\mathrm{eff}}$,
we obtain two types of vacuum diagrams in Figure\,\ref{fig:ge}.
Here, double wiggly and solid lines denote $h_{MN}$ and $\psi$, respectively.
Note that the figure-eight diagram (the left diagram in Figure\,\ref{fig:ge}) does not contribute to the effective potential
(see Appendix\,\ref{sec:ap3}).
Thus, we calculate only the right diagram in Figure\,\ref{fig:ge}:
\begin{align}
\label{5depge}
&-i\hat{V}_{\mathrm{Fig}.\,\ref{fig:ge}}^{2\mathchar`-\mathrm{loop}} \nonumber \\
&=\frac{(i\kappa)^2}{2!}\int \frac{d^5k}{(2\pi)^5}\int \frac{d^5p}{(2\pi)^5}\int \frac{d^5q}{(2\pi)^5}
(2\pi)^5\delta^{(5)}(\hat{k}+\hat{p}-\hat{q}) \nonumber \\
&~~~~~
\times\frac{1}{16}\bigl\{-\tilde{D}_{MNM'N'}(\hat{k})+\eta_{M'N'}\eta^{P'Q'}\tilde{D}_{MNP'Q'}(\hat{k}) \nonumber \\
&~~~~~~~~~~~~~~
+\eta_{MN}\eta^{PQ}\tilde{D}_{PQM'N'}(\hat{k})
-\eta_{MN}\eta_{M'N'}\eta^{PQ}\eta^{P'Q'}\tilde{D}_{PQP'Q'}(\hat{k})\bigr\} \nonumber \\
&~~~~~
\times\mathrm{tr}\bigl\{\Gamma^M\tilde{S}^N(\hat{\tilde{p}})\Gamma^{M'}\tilde{S}^{N'}(\hat{\tilde{q}})
+\Gamma^M\tilde{S}^{N'}(\hat{\tilde{p}})\Gamma^{M'}\tilde{S}^N(\hat{\tilde{q}}) \nonumber \\
&~~~~~~~~~~~~~~
+\Gamma^M\tilde{S}^{NN'}(\hat{\tilde{p}})\Gamma^{M'}\tilde{S}(\hat{\tilde{q}})
+\Gamma^M\tilde{S}(\hat{\tilde{p}})\Gamma^{M'}\tilde{S}^{N'N}(\hat{\tilde{q}})\bigr\} \nonumber \\
&=i\kappa^2\int \frac{d^5k_{\mathrm{E}}}{(2\pi)^5}\int \frac{d^5p_{\mathrm{E}}}{(2\pi)^5}
\biggl(\frac{10}{3}\frac{1}{\hat{k}_{\mathrm{E}}^2}
+\frac{29}{12}\frac{\hat{k}_{\mathrm{E}}\hat{\tilde{p}}_{\mathrm{E}}}{\hat{k}_{\mathrm{E}}^2\hat{\tilde{p}}_{\mathrm{E}}^2}\biggr)
-\frac{3}{8}i\kappa^2\int \frac{d^5p_{\mathrm{E}}}{(2\pi)^5}\int \frac{d^5q_{\mathrm{E}}}{(2\pi)^5}
\frac{\hat{\tilde{p}}_{\mathrm{E}}\hat{\tilde{q}}_{\mathrm{E}}}{\hat{\tilde{p}}_{\mathrm{E}}^2\hat{\tilde{q}}_{\mathrm{E}}^2}
\nonumber \\
&=\frac{i\kappa^2}{(2\pi R)^2}\sum_{\ell=-\infty}^{\infty}\sum_{m=-\infty}^{\infty}
\int \frac{d^4k_{\mathrm{E}}}{(2\pi)^4}\int \frac{d^4p_{\mathrm{E}}}{(2\pi)^4}
\biggl(\frac{10}{3}\frac{1}{\{k_{\mathrm{E}}^2+(\frac{\ell}{R})^2\}}
+\frac{29}{12}\frac{k_{\mathrm{E}}p_{\mathrm{E}}+\frac{\ell}{R}(\frac{m}{R}-\frac{\theta}{2\pi R})}
{\{k_{\mathrm{E}}^2+(\frac{\ell}{R})^2\}\{p_{\mathrm{E}}^2+(\frac{m}{R}-\frac{\theta}{2\pi R})^2\}}\biggr)
 \nonumber \\
&~~~~~~
-\frac{3}{8}\frac{i\kappa^2}{(2\pi R)^2}\sum_{m=-\infty}^{\infty}\sum_{n=-\infty}^{\infty}
\int \frac{d^4p_{\mathrm{E}}}{(2\pi)^4}\int \frac{d^4q_{\mathrm{E}}}{(2\pi)^4}
\frac{p_{\mathrm{E}}q_{\mathrm{E}}+(\frac{m}{R}-\frac{\theta}{2\pi R})(\frac{n}{R}-\frac{\theta}{2\pi R})}
{\{p_{\mathrm{E}}^2+(\frac{m}{R}-\frac{\theta}{2\pi R})^2\}\{q_{\mathrm{E}}^2+(\frac{n}{R}-\frac{\theta}{2\pi R})^2\}},
\end{align}
where $\tilde{D}_{MNPQ}(\hat{k})$, $\tilde{S}^M(\hat{\tilde{p}})$, and $\tilde{S}^{MN}(\hat{\tilde{p}})$
are defined in Appendix\,\ref{sec:ap1}.
In the expression (\ref{5depge}), odd functions for $k_{\mathrm{E}}$, $p_{\mathrm{E}}$, $q_{\mathrm{E}}$, and $\ell$ vanish
after the integrations are carried out, and we can rewrite the infinite sums using the formula (\ref{tif2}).
Consequently, the contribution to the 4D two-loop effective potential is expressed as
\begin{align}
\label{4d2pge}
V_{\mathrm{Fig}.\,\ref{fig:ge}}^{2\mathchar`-\mathrm{loop}}(\theta)
=\frac{3}{8}\frac{3^2\kappa_4^2}{16\pi^4(2\pi R)^6}
\sum_{m,n=1}^{\infty}\frac{\sin(m\theta)}{m^4}\frac{\sin(n\theta)}{n^4},
\end{align}
where $V_{\mathrm{Fig}.\,\ref{fig:ge}}^{2\mathchar`-\mathrm{loop}}
=2\pi R\hat{V}_{\mathrm{Fig}.\,\ref{fig:ge}}^{2\mathchar`-\mathrm{loop}}$,
$\kappa_4(=\kappa/\sqrt{2\pi R})$ is the 4D gravitational coupling constant,
and terms with $\theta$-independent divergences are dropped.

It is shown that the expression (\ref{4d2pge}) is the finite periodic function for $\theta$.
The result is consistent with the conjecture that the effective potential for the Higgs boson is finite owing to
the higher-dimensional gauge symmetry.
In this case, its minimum is localized at $\theta=0$ unlike an $U(1)$ gauge theory without graviton in Sect.\,2.
However, the vacuum remains at $\theta=\pi$, since $V_{\mathrm{Fig}.\,\ref{fig:ge}}^{2\mathchar`-\mathrm{loop}}$
is much smaller than $V^{1\mathchar`-\mathrm{loop}}$ and $V_{\mathrm{Fig}.\,\ref{fig:pe}}^{2\mathchar`-\mathrm{loop}}$
because of the tininess of the gravitational coupling constant.
Therefore, the expression (\ref{4d2pge}) does not contribute to phenomenological analyses.

\section{Conclusions and discussions}
\label{Cad}
We have investigated the two-loop finiteness of the effective potential for the Higgs boson in the $U(1)$ GHU
model coupled to QG on $M^4\times S^1$.
Although the potential is expected to be finite due to gauge invariance, there is no proof to guarantee the finiteness.
In both Abelian and non-Abelian gauge theories without graviton, the finiteness has been verified at the two-loop level
\cite{Shiraishi:1991hs}\cite{Maru:2006wa}\cite{Hosotani:2007kn}\cite{Hisano:2019cxm}.
We have found the two-loop finiteness in the case with graviton at the first time.

Since interactions between gravitons and fermions are three and four-points at the tree level,
there are two types of diagrams including quantum gravitational corrections (see Appendix\,\ref{sec:ap3}).
We have found that one diagram is irrelevant to the potential for the Higgs boson $\theta$
while another diagram gives the finite periodic function for $\theta$ indeed.
Hence, it has turned out that the graviton contributions to the effective potential is finite at the two-loop level.

In our result, although the function is minimized at $\theta=\pi$ unlike the Abelian case without graviton,
it does not affect phenomenological analyses because of the tininess of the gravitational coupling constant.
Nevertheless, our result has an important implication.
If the finiteness were damaged by graviton, the GHU models would no longer be a good candidate
of new physics compatible with QG.
In other words, the GHU remains a powerful concept by virtue of finiteness even if the models are coupled to QG.
Our result holds true in an extranatural inflation model with a radion analyzed in \cite{Abe:2014eia}\cite{Abe:2015bba}.

We have supposed a circle $S^1$ as an extra-space for simplicity.
It is well known that the compactification on the orbifold $S^1/Z_2$ is useful
to derive chiral fermions after dimensional reduction.
We can take not only a flat metric but also the Randall-Sundrum (RS) metric as background including the orbifold.
The RS metric has been applied for constructing realistic GHU models with the characteristics of the SM
\cite{Agashe:2004rs}\cite{Hosotani:2008tx}\cite{Funatsu:2013ni}.
On the RS background, effects from the graviton localized around the UV brane are expected to be negligibly small
because they are exponentially suppressed.
Nevertheless, it is still important to study the finiteness of the effective potential.

\section*{Acknowledgments}
The author thanks Prof.\,Kawamura for valuable discussions.

\appendix
\section{Propagators}
\label{sec:ap1}
We define functions concerning propagators as
\begin{align}
D_{MN}(\hat{x}_1,\hat{x}_2)&=\int\frac{d^5k}{(2\pi)^5}e^{-i\hat{k}(\hat{x}_1-\hat{x}_2)}\tilde{D}_{MN}(\hat{k}), \nonumber \\
D_{MNPQ}(\hat{x}_1,\hat{x}_2)&=\int\frac{d^5k}{(2\pi)^5}e^{-i\hat{k}(\hat{x}_1-\hat{x}_2)}\tilde{D}_{MNPQ}(\hat{k}), \nonumber \\
S(\hat{x}_1,\hat{x}_2)&=\int\frac{d^5p}{(2\pi)^5}e^{-i\hat{p}(\hat{x}_1-\hat{x}_2)}\tilde{S}(\hat{\tilde{p}}), \nonumber \\
S^M(\hat{x}_1,\hat{x}_2)&=\int\frac{d^5p}{(2\pi)^5}e^{-i\hat{p}(\hat{x}_1-\hat{x}_2)}\tilde{S}^{M}(\hat{\tilde{p}}), \nonumber \\
S^{MN}(\hat{x}_1,\hat{x}_2)&=\int\frac{d^5p}{(2\pi)^5}e^{-i\hat{p}(\hat{x}_1-\hat{x}_2)}\tilde{S}^{MN}(\hat{\tilde{p}}),
\end{align}
where integrands are given by
\begin{align}
\label{pro}
\tilde{D}_{MN}(\hat{k})&\equiv\frac{-i\eta_{MN}}{\hat{k}^2+i\epsilon},~~
\tilde{D}_{MNPQ}(\hat{k})\equiv\frac{i\mathscr{D}_{MNPQ}}{\hat{k}^2+i\epsilon},~~
\tilde{S}(\hat{\tilde{p}})\equiv\frac{i}{\fsl{\hat{\tilde{p}}}+i\epsilon}, \nonumber \\
&~~~~~
\tilde{S}^M(\hat{\tilde{p}})\equiv\frac{i\tilde{p}^M}{\fsl{\hat{\tilde{p}}}+i\epsilon},~~
\tilde{S}^{MN}(\hat{\tilde{p}})\equiv\frac{i\tilde{p}^M\tilde{p}^N}{\fsl{\hat{\tilde{p}}}+i\epsilon},
\end{align}
with
\begin{align}
\label{cogp}
\mathscr{D}_{MNPQ}=\eta_{MP}\eta_{NQ}+\eta_{MQ}\eta_{NP}-\frac{2}{3}\eta_{MN}\eta_{PQ}.
\end{align}
Here, $D_{MN}(\hat{x}_1,\hat{x}_2)$, $D_{MNPQ}(\hat{x}_1,\hat{x}_2)$,
and $S(\hat{x}_1,\hat{x}_2)$ represent propagators for
$U(1)$ gauge boson, graviton, and fermion.

\section{Formulas for infinite summations}
\label{sec:ap2}
We write down several formulas relating to the effective potential.
First formula is given as
\begin{align}
\label{tif1}
\frac{1}{2\pi R}\sum_{m=-\infty}^{\infty}
\int \frac{d^4p_{\mathrm{E}}}{(2\pi)^4}
\frac{1}{p_{\mathrm{E}}^2+(\frac{m}{R}-\frac{\theta}{2\pi R})^2}
=\frac{1}{4\pi^2(2\pi R)^3}\sum_{m=1}^{\infty}\frac{\cos(m\theta)}{m^3},
\end{align}
where terms with a $\theta$-independent divergence are dropped.

Next one is given as
\begin{align}
\label{tif2}
\frac{1}{2\pi R}\sum_{m=-\infty}^{\infty}
\int \frac{d^4p_{\mathrm{E}}}{(2\pi)^4}
\frac{\frac{m}{R}-\frac{\theta}{2\pi R}}
{p_{\mathrm{E}}^2+(\frac{m}{R}-\frac{\theta}{2\pi R})^2}
&=\frac{1}{32\pi^3R^2}\int_0^\infty ds~s^{-2}
\sum_{m=-\infty}^{\infty}
\biggl(m-\frac{\theta}{2\pi}\biggr)
e^{-s(\frac{m}{R}-\frac{\theta}{2\pi R})^2} \nonumber \\
&=-\frac{\sqrt{\pi}R}{16\pi^2}\sum_{m=1}^{\infty}m\sin(m\theta)
\int_0^\infty ds~s^{-\frac{7}{2}}e^{-\frac{\pi^2R^2m^2}{s}} \nonumber \\
&=-\frac{3}{4\pi^2(2\pi R)^4}\sum_{m=1}^{\infty}\frac{\sin(m\theta)}{m^4},
\end{align}
where the following Poisson resummation formulas are used
\begin{align}
\sum_{m=-\infty}^{\infty}e^{-s(\frac{m}{R}-\frac{\theta}{2\pi R})^2}
&=R\sqrt{\frac{\pi}{s}}\biggl\{1+2\sum_{m=1}^{\infty}e^{-\frac{\pi^2R^2m^2}{s}}\cos(m\theta)\biggr\}, \nonumber \\
\sum_{m=-\infty}^{\infty}me^{-s(\frac{m}{R}-\frac{\theta}{2\pi R})^2}
&=R\sqrt{\frac{\pi}{s}}\biggl[\frac{\theta}{2\pi}+2\sum_{m=1}^{\infty}e^{-\frac{\pi^2R^2m^2}{s}}
\biggl\{\frac{\theta}{2\pi}\cos(m\theta)-\frac{\pi R^2}{s}m\sin(m\theta)\biggr\}\biggr].
\end{align}

\section{Two-loop calculation}
\label{sec:ap3}
We can expand the Lagrangian density (\ref{5dlge})
using the relations $g_{MN}=\eta_{MN}+\kappa h_{MN}$, (\ref{me}), and (\ref{ve}).
Thus, we obtain the interaction terms between gravitons and fermions\cite{Donoghue:2015hwa}:
\begin{align}
\label{5del12}
\hat{\mathcal{L}}_{\psi}^{(1)}
&=-\frac{1}{4}\kappa h_{MN}\bar{\psi}i\Gamma^M\tilde{\partial}^N\psi
+\frac{1}{4}\kappa h_{MN}\tilde{\partial}^{*N}\bar{\psi}i\Gamma^M\psi \nonumber \\
&~~~~
+\frac{1}{4}\kappa \eta_{MN}h\bar{\psi}i\Gamma^M\tilde{\partial}^N\psi
-\frac{1}{4}\kappa \eta_{MN}h\tilde{\partial}^{*N}\bar{\psi}i\Gamma^M\psi, \nonumber \\
\hat{\mathcal{L}}_{\psi}^{(2)}
&=\frac{3}{16}\kappa^2h_{MP}h_N^{\phantom{N}P}\bar{\psi}i\Gamma^M\tilde{\partial}^N\psi
-\frac{3}{16}\kappa^2h_{MP}h_N^{\phantom{N}P}\tilde{\partial}^{*N}\bar{\psi}i\Gamma^M\psi \nonumber \\
&~~~~
-\frac{1}{8}\kappa^2h_{MN}h\bar{\psi}i\Gamma^M\tilde{\partial}^N\psi
+\frac{1}{8}\kappa^2h_{MN}h\tilde{\partial}^{*N}\bar{\psi}i\Gamma^M\psi \nonumber \\
&~~~~
-\frac{1}{8}\kappa^2\eta_{MN}h_{PQ}h^{PQ}\bar{\psi}i\Gamma^M\tilde{\partial}^N\psi
+\frac{1}{8}\kappa^2\eta_{MN}h_{PQ}h^{PQ}\tilde{\partial}^{*N}\bar{\psi}i\Gamma^M\psi \nonumber \\
&~~~~
+\frac{1}{16}\kappa^2\eta_{MN}h^2\bar{\psi}i\Gamma^M\tilde{\partial}^N\psi
-\frac{1}{16}\kappa^2\eta_{MN}h^2\tilde{\partial}^{*N}\bar{\psi}i\Gamma^M\psi \nonumber \\
&~~~~
+\frac{1}{32}\kappa^2\partial_Mh_{PS}h_Q^{\phantom{Q}S}\bar{\psi}i\bigl\{\Gamma^M,\Gamma^{PQ}\bigr\}\psi,
\end{align}
where $\hat{\mathcal{L}}_{\psi}^{(1)}$ and $\hat{\mathcal{L}}_{\psi}^{(2)}$ are quantities of $\mathcal{O}(h)$
and $\mathcal{O}(h^2)$, respectively.
These terms contribute to the two-loop effective potential\cite{Jackiw:1974cv}\cite{Iliopoulos:1974ur}.

From the interactions in $\hat{\mathcal{L}}_{\psi}^{(2)}$,
the process represented by the left diagram in Figure\,\ref{fig:ge} occurs.
Although it involves fermions, it does not contribute to the effective potential for $\theta$.
Indeed, it is expressed as
\begin{align}
\label{5d2epe}
&i\kappa^2\int \frac{d^5k_{\mathrm{E}}}{(2\pi)^5}\int \frac{d^5p_{\mathrm{E}}}{(2\pi)^5}
\biggl(c_1\frac{\hat{\tilde{p}}_{\mathrm{E}}^2}{\hat{k}_{\mathrm{E}}^2\hat{\tilde{p}}_{\mathrm{E}}^2}
+c_2\frac{\hat{k}_{\mathrm{E}}\hat{\tilde{p}}_{\mathrm{E}}}{\hat{k}_{\mathrm{E}}^2\hat{\tilde{p}}_{\mathrm{E}}^2}\biggr)
\nonumber \\
&=\frac{i\kappa^2}{(2\pi R)^2}\sum_{\ell=-\infty}^{\infty}\sum_{m=-\infty}^{\infty}
\int \frac{d^4k_{\mathrm{E}}}{(2\pi)^4}\int \frac{d^4p_{\mathrm{E}}}{(2\pi)^4}
\biggl(c_1\frac{1}{k_{\mathrm{E}}^2+(\frac{\ell}{R})^2}
+c_2\frac{k_{\mathrm{E}}p_{\mathrm{E}}+\frac{\ell}{R}(\frac{m}{R}-\frac{\theta}{2\pi R})}
{\{k_{\mathrm{E}}^2+(\frac{\ell}{R})^2\}\{p_{\mathrm{E}}^2+(\frac{m}{R}-\frac{\theta}{2\pi R})^2\}}\biggr),
\end{align}
where $c_1$ and $c_2$ are constants.
The first term does not contain $\theta$,
and other terms equal zero because they are odd functions for $k_{\mathrm{E}}$, $p_{\mathrm{E}}$, and $\ell$.

On the other hand, from the interactions in $\hat{\mathcal{L}}_{\psi}^{(1)}$,
the process represented by the right diagram in Figure\,\ref{fig:ge} occurs.
Only this diagram contributes to the two-loop effective potential:
\begin{align}
\label{5depc}
&i\int d^5x\bigl(-\hat{V}_{\mathrm{Fig}.\,\ref{fig:ge}}^{2\mathchar`-\mathrm{loop}}\bigr) \nonumber \\
&=
\textstyle
\contraction[1ex]{\bigl\langle{\frac{1}{2!}}i\int d^5x_1\bigl(-\frac{1}{4}\kappa}{h}
{{}_{MN}\bar{\psi}i\Gamma^M\tilde{\partial}^N\psi\bigr)i\int d^5x_2\bigl(-\frac{1}{4}\kappa}{h}
\contraction[1.5ex]{\bigl\langle{\frac{1}{2!}}i\int d^5x_1\bigl(-\frac{1}{4}\kappa h_{MN}}{\bar{\psi}}
{i\Gamma^M\tilde{\partial}^N\psi\bigr)i\int d^5x_2\bigl(-\frac{1}{4}\kappa h_{M'N'}\bar{\psi}i\Gamma^{M'}\tilde{\partial}^{N'}}{\psi}
\contraction[2.6ex]{\bigl\langle{\frac{1}{2!}}i\int d^5x_1\bigl(-\frac{1}{4}\kappa h_{MN}\bar{\psi}i\Gamma^M\tilde{\partial}^N}{\psi}
{\bigr)i\int d^5x_2\bigl(-\frac{1}{4}\kappa h_{M'N'}}{\bar{\psi}}
\bigl\langle{\frac{1}{2!}}
i\int d^5x_1\bigl(-\frac{1}{4}\kappa h_{MN}\bar{\psi}i\Gamma^M\tilde{\partial}^N\psi\bigr)
i\int d^5x_2\bigl(-\frac{1}{4}\kappa h_{M'N'}\bar{\psi}i\Gamma^{M'}\tilde{\partial}^{N'}\psi\bigr)
\bigr\rangle \nonumber \\
&~~~~+
\textstyle
\contraction[1ex]{\bigl\langle{\frac{1}{2!}}i\int d^5x_1\bigl(-\frac{1}{4}\kappa}{h}
{{}_{MN}\bar{\psi}i\Gamma^M\tilde{\partial}^N\psi\bigr)i\int d^5x_2\bigl(\frac{1}{4}\kappa}{h}
\contraction[1.5ex]{\bigl\langle{\frac{1}{2!}}i\int d^5x_1\bigl(-\frac{1}{4}\kappa h_{MN}}{\bar{\psi}}
{i\Gamma^M\tilde{\partial}^N\psi\bigr)i\int d^5x_2\bigl(\frac{1}{4}\kappa h_{M'N'}\tilde{\partial}^{*N'}\bar{\psi}i\Gamma^{M'}}{\psi}
\contraction[2.6ex]{\bigl\langle{\frac{1}{2!}}i\int d^5x_1\bigl(-\frac{1}{4}\kappa h_{MN}\bar{\psi}i\Gamma^M\tilde{\partial}^N}{\psi}
{\bigr)i\int d^5x_2\bigl(\frac{1}{4}\kappa h_{M'N'}\tilde{\partial}^{*N'}}{\bar{\psi}}
\bigl\langle{\frac{1}{2!}}
i\int d^5x_1\bigl(-\frac{1}{4}\kappa h_{MN}\bar{\psi}i\Gamma^M\tilde{\partial}^N\psi\bigr)
i\int d^5x_2\bigl(\frac{1}{4}\kappa h_{M'N'}\tilde{\partial}^{*N'}\bar{\psi}i\Gamma^{M'}\psi\bigr)
\bigr\rangle \nonumber \\
&~~~~+
\textstyle
\contraction[1ex]{\bigl\langle{\frac{1}{2!}}i\int d^5x_1\bigl(-\frac{1}{4}\kappa}{h}
{{}_{MN}\bar{\psi}i\Gamma^M\tilde{\partial}^N\psi\bigr)i\int d^5x_2\bigl(\frac{1}{4}\kappa\eta_{M'N'}}{h}
\contraction[1.5ex]{\bigl\langle{\frac{1}{2!}}i\int d^5x_1\bigl(-\frac{1}{4}\kappa h_{MN}}{\bar{\psi}}
{i\Gamma^M\tilde{\partial}^N\psi\bigr)i\int d^5x_2\bigl(\frac{1}{4}\kappa\eta_{M'N'}h\bar{\psi}i\Gamma^{M'}\tilde{\partial}^{N'}}{\psi}
\contraction[2.6ex]{\bigl\langle{\frac{1}{2!}}i\int d^5x_1\bigl(-\frac{1}{4}\kappa h_{MN}\bar{\psi}i\Gamma^M\tilde{\partial}^N}{\psi}
{\bigr)i\int d^5x_2\bigl(\frac{1}{4}\kappa\eta_{M'N'}h}{\bar{\psi}}
\bigl\langle{\frac{1}{2!}}
i\int d^5x_1\bigl(-\frac{1}{4}\kappa h_{MN}\bar{\psi}i\Gamma^M\tilde{\partial}^N\psi\bigr)
i\int d^5x_2\bigl(\frac{1}{4}\kappa\eta_{M'N'}h\bar{\psi}i\Gamma^{M'}\tilde{\partial}^{N'}\psi\bigr)
\bigr\rangle \nonumber \\
&~~~~+
\textstyle
\contraction[1ex]{\bigl\langle{\frac{1}{2!}}i\int d^5x_1\bigl(-\frac{1}{4}\kappa}{h}
{{}_{MN}\bar{\psi}i\Gamma^M\tilde{\partial}^N\psi\bigr)i\int d^5x_2\bigl(-\frac{1}{4}\kappa\eta_{M'N'}}{h}
\contraction[1.5ex]{\bigl\langle{\frac{1}{2!}}i\int d^5x_1\bigl(-\frac{1}{4}\kappa h_{MN}}{\bar{\psi}}
{i\Gamma^M\tilde{\partial}^N\psi\bigr)i\int d^5x_2\bigl(-\frac{1}{4}\kappa\eta_{M'N'}h\tilde{\partial}^{*N'}\bar{\psi}i\Gamma^{M'}}{\psi}
\contraction[2.6ex]{\bigl\langle{\frac{1}{2!}}i\int d^5x_1\bigl(-\frac{1}{4}\kappa h_{MN}\bar{\psi}i\Gamma^M\tilde{\partial}^N}{\psi}
{\bigr)i\int d^5x_2\bigl(-\frac{1}{4}\kappa\eta_{M'N'}h\tilde{\partial}^{*N'}}{\bar{\psi}}
\bigl\langle{\frac{1}{2!}}
i\int d^5x_1\bigl(-\frac{1}{4}\kappa h_{MN}\bar{\psi}i\Gamma^M\tilde{\partial}^N\psi\bigr)
i\int d^5x_2\bigl(-\frac{1}{4}\kappa\eta_{M'N'}h\tilde{\partial}^{*N'}\bar{\psi}i\Gamma^{M'}\psi\bigr)
\bigr\rangle \nonumber \\
&~~~~+
\textstyle
\contraction[1ex]{\bigl\langle{\frac{1}{2!}}i\int d^5x_1\bigl(\frac{1}{4}\kappa}{h}
{{}_{MN}\tilde{\partial}^{*N}\bar{\psi}i\Gamma^M\psi\bigr)i\int d^5x_2\bigl(-\frac{1}{4}\kappa}{h}
\contraction[1.5ex]{\bigl\langle{\frac{1}{2!}}i\int d^5x_1\bigl(\frac{1}{4}\kappa h_{MN}\tilde{\partial}^{*N}}{\bar{\psi}}
{i\Gamma^M\psi\bigr)i\int d^5x_2\bigl(-\frac{1}{4}\kappa h_{M'N'}\bar{\psi}i\Gamma^{M'}\tilde{\partial}^{N'}}{\psi}
\contraction[2.6ex]{\bigl\langle{\frac{1}{2!}}i\int d^5x_1\bigl(\frac{1}{4}\kappa h_{MN}\tilde{\partial}^{*N}\bar{\psi}i\Gamma^M}{\psi}
{\bigr)i\int d^5x_2\bigl(-\frac{1}{4}\kappa h_{M'N'}}{\bar{\psi}}
\bigl\langle{\frac{1}{2!}}
i\int d^5x_1\bigl(\frac{1}{4}\kappa h_{MN}\tilde{\partial}^{*N}\bar{\psi}i\Gamma^M\psi\bigr)
i\int d^5x_2\bigl(-\frac{1}{4}\kappa h_{M'N'}\bar{\psi}i\Gamma^{M'}\tilde{\partial}^{N'}\psi\bigr)
\bigr\rangle \nonumber \\
&~~~~+
\textstyle
\contraction[1ex]{\bigl\langle{\frac{1}{2!}}i\int d^5x_1\bigl(\frac{1}{4}\kappa}{h}
{{}_{MN}\tilde{\partial}^{*N}\bar{\psi}i\Gamma^M\psi\bigr)i\int d^5x_2\bigl(\frac{1}{4}\kappa}{h}
\contraction[1.5ex]{\bigl\langle{\frac{1}{2!}}i\int d^5x_1\bigl(\frac{1}{4}\kappa h_{MN}\tilde{\partial}^{*N}}{\bar{\psi}}
{i\Gamma^M\psi\bigr)i\int d^5x_2\bigl(\frac{1}{4}\kappa h_{M'N'}\tilde{\partial}^{*N'}\bar{\psi}i\Gamma^{M'}}{\psi}
\contraction[2.6ex]{\bigl\langle{\frac{1}{2!}}i\int d^5x_1\bigl(\frac{1}{4}\kappa h_{MN}\tilde{\partial}^{*N}\bar{\psi}i\Gamma^M}{\psi}
{\bigr)i\int d^5x_2\bigl(\frac{1}{4}\kappa h_{M'N'}\tilde{\partial}^{*N'}}{\bar{\psi}}
\bigl\langle{\frac{1}{2!}}
i\int d^5x_1\bigl(\frac{1}{4}\kappa h_{MN}\tilde{\partial}^{*N}\bar{\psi}i\Gamma^M\psi\bigr)
i\int d^5x_2\bigl(\frac{1}{4}\kappa h_{M'N'}\tilde{\partial}^{*N'}\bar{\psi}i\Gamma^{M'}\psi\bigr)
\bigr\rangle \nonumber \\
&~~~~+
\textstyle
\contraction[1ex]{\bigl\langle{\frac{1}{2!}}i\int d^5x_1\bigl(\frac{1}{4}\kappa}{h}
{{}_{MN}\tilde{\partial}^{*N}\bar{\psi}i\Gamma^M\psi\bigr)i\int d^5x_2\bigl(\frac{1}{4}\kappa\eta_{M'N'}}{h}
\contraction[1.5ex]{\bigl\langle{\frac{1}{2!}}i\int d^5x_1\bigl(\frac{1}{4}\kappa h_{MN}\tilde{\partial}^{*N}}{\bar{\psi}}
{i\Gamma^M\psi\bigr)i\int d^5x_2\bigl(\frac{1}{4}\kappa\eta_{M'N'}h\bar{\psi}i\Gamma^{M'}\tilde{\partial}^{N'}}{\psi}
\contraction[2.6ex]{\bigl\langle{\frac{1}{2!}}i\int d^5x_1\bigl(\frac{1}{4}\kappa h_{MN}\tilde{\partial}^{*N}\bar{\psi}i\Gamma^M}{\psi}
{\bigr)i\int d^5x_2\bigl(\frac{1}{4}\kappa\eta_{M'N'}h}{\bar{\psi}}
\bigl\langle{\frac{1}{2!}}
i\int d^5x_1\bigl(\frac{1}{4}\kappa h_{MN}\tilde{\partial}^{*N}\bar{\psi}i\Gamma^M\psi\bigr)
i\int d^5x_2\bigl(\frac{1}{4}\kappa\eta_{M'N'}h\bar{\psi}i\Gamma^{M'}\tilde{\partial}^{N'}\psi\bigr)
\bigr\rangle \nonumber \\
&~~~~+
\textstyle
\contraction[1ex]{\bigl\langle{\frac{1}{2!}}i\int d^5x_1\bigl(\frac{1}{4}\kappa}{h}
{{}_{MN}\tilde{\partial}^{*N}\bar{\psi}i\Gamma^M\psi\bigr)i\int d^5x_2\bigl(-\frac{1}{4}\kappa\eta_{M'N'}}{h}
\contraction[1.5ex]{\bigl\langle{\frac{1}{2!}}i\int d^5x_1\bigl(\frac{1}{4}\kappa h_{MN}\tilde{\partial}^{*N}}{\bar{\psi}}
{i\Gamma^M\psi\bigr)i\int d^5x_2\bigl(-\frac{1}{4}\kappa\eta_{M'N'}h\tilde{\partial}^{*N'}\bar{\psi}i\Gamma^{M'}}{\psi}
\contraction[2.6ex]{\bigl\langle{\frac{1}{2!}}i\int d^5x_1\bigl(\frac{1}{4}\kappa h_{MN}\tilde{\partial}^{*N}\bar{\psi}i\Gamma^M}{\psi}
{\bigr)i\int d^5x_2\bigl(-\frac{1}{4}\kappa\eta_{M'N'}h\tilde{\partial}^{*N'}}{\bar{\psi}}
\bigl\langle{\frac{1}{2!}}
i\int d^5x_1\bigl(\frac{1}{4}\kappa h_{MN}\tilde{\partial}^{*N}\bar{\psi}i\Gamma^M\psi\bigr)
i\int d^5x_2\bigl(-\frac{1}{4}\kappa\eta_{M'N'}h\tilde{\partial}^{*N'}\bar{\psi}i\Gamma^{M'}\psi\bigr)
\bigr\rangle \nonumber \\
&~~~~+
\textstyle
\contraction[1ex]{\bigl\langle{\frac{1}{2!}}i\int d^5x_1\bigl(\frac{1}{4}\kappa\eta_{MN}}{h}
{\bar{\psi}i\Gamma^M\tilde{\partial}^N\psi\bigr)i\int d^5x_2\bigl(-\frac{1}{4}\kappa}{h}
\contraction[1.5ex]{\bigl\langle{\frac{1}{2!}}i\int d^5x_1\bigl(\frac{1}{4}\kappa\eta_{MN}h}{\bar{\psi}}
{i\Gamma^M\tilde{\partial}^N\psi\bigr)i\int d^5x_2\bigl(-\frac{1}{4}\kappa h_{M'N'}\bar{\psi}i\Gamma^{M'}\tilde{\partial}^{N'}}{\psi}
\contraction[2.6ex]{\bigl\langle{\frac{1}{2!}}i\int d^5x_1\bigl(\frac{1}{4}\kappa\eta_{MN}h\bar{\psi}i\Gamma^M\tilde{\partial}^N}{\psi}
{\bigr)i\int d^5x_2\bigl(-\frac{1}{4}\kappa h_{M'N'}}{\bar{\psi}}
\bigl\langle{\frac{1}{2!}}
i\int d^5x_1\bigl(\frac{1}{4}\kappa\eta_{MN}h\bar{\psi}i\Gamma^M\tilde{\partial}^N\psi\bigr)
i\int d^5x_2\bigl(-\frac{1}{4}\kappa h_{M'N'}\bar{\psi}i\Gamma^{M'}\tilde{\partial}^{N'}\psi\bigr)
\bigr\rangle \nonumber \\
&~~~~+
\textstyle
\contraction[1ex]{\bigl\langle{\frac{1}{2!}}i\int d^5x_1\bigl(\frac{1}{4}\kappa\eta_{MN}}{h}
{\bar{\psi}i\Gamma^M\tilde{\partial}^N\psi\bigr)i\int d^5x_2\bigl(\frac{1}{4}\kappa}{h}
\contraction[1.5ex]{\bigl\langle{\frac{1}{2!}}i\int d^5x_1\bigl(\frac{1}{4}\kappa\eta_{MN}h}{\bar{\psi}}
{i\Gamma^M\tilde{\partial}^N\psi\bigr)i\int d^5x_2\bigl(\frac{1}{4}\kappa h_{M'N'}\tilde{\partial}^{*N'}\bar{\psi}i\Gamma^{M'}}{\psi}
\contraction[2.6ex]{\bigl\langle{\frac{1}{2!}}i\int d^5x_1\bigl(\frac{1}{4}\kappa\eta_{MN}h\bar{\psi}i\Gamma^M\tilde{\partial}^N}{\psi}
{\bigr)i\int d^5x_2\bigl(\frac{1}{4}\kappa h_{M'N'}\tilde{\partial}^{*N'}}{\bar{\psi}}
\bigl\langle{\frac{1}{2!}}
i\int d^5x_1\bigl(\frac{1}{4}\kappa\eta_{MN}h\bar{\psi}i\Gamma^M\tilde{\partial}^N\psi\bigr)
i\int d^5x_2\bigl(\frac{1}{4}\kappa h_{M'N'}\tilde{\partial}^{*N'}\bar{\psi}i\Gamma^{M'}\psi\bigr)
\bigr\rangle \nonumber \\
&~~~~+
\textstyle
\contraction[1ex]{\bigl\langle{\frac{1}{2!}}i\int d^5x_1\bigl(\frac{1}{4}\kappa\eta_{MN}}{h}
{\bar{\psi}i\Gamma^M\tilde{\partial}^N\psi\bigr)i\int d^5x_2\bigl(\frac{1}{4}\kappa\eta_{M'N'}}{h}
\contraction[1.5ex]{\bigl\langle{\frac{1}{2!}}i\int d^5x_1\bigl(\frac{1}{4}\kappa\eta_{MN}h}{\bar{\psi}}
{i\Gamma^M\tilde{\partial}^N\psi\bigr)i\int d^5x_2\bigl(\frac{1}{4}\kappa\eta_{M'N'}h\bar{\psi}i\Gamma^{M'}\tilde{\partial}^{N'}}{\psi}
\contraction[2.6ex]{\bigl\langle{\frac{1}{2!}}i\int d^5x_1\bigl(\frac{1}{4}\kappa\eta_{MN}h\bar{\psi}i\Gamma^M\tilde{\partial}^N}{\psi}
{\bigr)i\int d^5x_2\bigl(\frac{1}{4}\kappa\eta_{M'N'}h}{\bar{\psi}}
\bigl\langle{\frac{1}{2!}}
i\int d^5x_1\bigl(\frac{1}{4}\kappa\eta_{MN}h\bar{\psi}i\Gamma^M\tilde{\partial}^N\psi\bigr)
i\int d^5x_2\bigl(\frac{1}{4}\kappa\eta_{M'N'}h\bar{\psi}i\Gamma^{M'}\tilde{\partial}^{N'}\psi\bigr)
\bigr\rangle \nonumber \\
&~~~~+
\textstyle
\contraction[1ex]{\bigl\langle{\frac{1}{2!}}i\int d^5x_1\bigl(\frac{1}{4}\kappa\eta_{MN}}{h}
{\bar{\psi}i\Gamma^M\tilde{\partial}^N\psi\bigr)i\int d^5x_2\bigl(-\frac{1}{4}\kappa\eta_{M'N'}}{h}
\contraction[1.5ex]{\bigl\langle{\frac{1}{2!}}i\int d^5x_1\bigl(\frac{1}{4}\kappa\eta_{MN}h}{\bar{\psi}}
{i\Gamma^M\tilde{\partial}^N\psi\bigr)i\int d^5x_2\bigl(-\frac{1}{4}\kappa\eta_{M'N'}h\tilde{\partial}^{*N'}\bar{\psi}i\Gamma^{M'}}{\psi}
\contraction[2.6ex]{\bigl\langle{\frac{1}{2!}}i\int d^5x_1\bigl(\frac{1}{4}\kappa\eta_{MN}h\bar{\psi}i\Gamma^M\tilde{\partial}^N}{\psi}
{\bigr)i\int d^5x_2\bigl(-\frac{1}{4}\kappa\eta_{M'N'}h\tilde{\partial}^{*N'}}{\bar{\psi}}
\bigl\langle{\frac{1}{2!}}
i\int d^5x_1\bigl(\frac{1}{4}\kappa\eta_{MN}h\bar{\psi}i\Gamma^M\tilde{\partial}^N\psi\bigr)
i\int d^5x_2\bigl(-\frac{1}{4}\kappa\eta_{M'N'}h\tilde{\partial}^{*N'}\bar{\psi}i\Gamma^{M'}\psi\bigr)
\bigr\rangle \nonumber \\
&~~~~+
\textstyle
\contraction[1ex]{\bigl\langle{\frac{1}{2!}}i\int d^5x_1\bigl(-\frac{1}{4}\kappa\eta_{MN}}{h}
{\tilde{\partial}^{*N}\bar{\psi}i\Gamma^M\psi\bigr)i\int d^5x_2\bigl(-\frac{1}{4}\kappa}{h}
\contraction[1.5ex]{\bigl\langle{\frac{1}{2!}}i\int d^5x_1\bigl(-\frac{1}{4}\kappa\eta_{MN}h\tilde{\partial}^{*N}}{\bar{\psi}}
{i\Gamma^M\psi\bigr)i\int d^5x_2\bigl(-\frac{1}{4}\kappa h_{M'N'}\bar{\psi}i\Gamma^{M'}\tilde{\partial}^{N'}}{\psi}
\contraction[2.6ex]{\bigl\langle{\frac{1}{2!}}i\int d^5x_1\bigl(-\frac{1}{4}\kappa\eta_{MN}h\tilde{\partial}^{*N}\bar{\psi}i\Gamma^M}{\psi}
{\bigr)i\int d^5x_2\bigl(-\frac{1}{4}\kappa h_{M'N'}}{\bar{\psi}}
\bigl\langle{\frac{1}{2!}}
i\int d^5x_1\bigl(-\frac{1}{4}\kappa\eta_{MN}h\tilde{\partial}^{*N}\bar{\psi}i\Gamma^M\psi\bigr)
i\int d^5x_2\bigl(-\frac{1}{4}\kappa h_{M'N'}\bar{\psi}i\Gamma^{M'}\tilde{\partial}^{N'}\psi\bigr)
\bigr\rangle \nonumber \\
&~~~~+
\textstyle
\contraction[1ex]{\bigl\langle{\frac{1}{2!}}i\int d^5x_1\bigl(-\frac{1}{4}\kappa\eta_{MN}}{h}
{\tilde{\partial}^{*N}\bar{\psi}i\Gamma^M\psi\bigr)i\int d^5x_2\bigl(\frac{1}{4}\kappa}{h}
\contraction[1.5ex]{\bigl\langle{\frac{1}{2!}}i\int d^5x_1\bigl(-\frac{1}{4}\kappa\eta_{MN}h\tilde{\partial}^{*N}}{\bar{\psi}}
{i\Gamma^M\psi\bigr)i\int d^5x_2\bigl(\frac{1}{4}\kappa h_{M'N'}\tilde{\partial}^{*N'}\bar{\psi}i\Gamma^{M'}}{\psi}
\contraction[2.6ex]{\bigl\langle{\frac{1}{2!}}i\int d^5x_1\bigl(-\frac{1}{4}\kappa\eta_{MN}h\tilde{\partial}^{*N}\bar{\psi}i\Gamma^M}{\psi}
{\bigr)i\int d^5x_2\bigl(\frac{1}{4}\kappa h_{M'N'}\tilde{\partial}^{*N'}}{\bar{\psi}}
\bigl\langle{\frac{1}{2!}}
i\int d^5x_1\bigl(-\frac{1}{4}\kappa\eta_{MN}h\tilde{\partial}^{*N}\bar{\psi}i\Gamma^M\psi\bigr)
i\int d^5x_2\bigl(\frac{1}{4}\kappa h_{M'N'}\tilde{\partial}^{*N'}\bar{\psi}i\Gamma^{M'}\psi\bigr)
\bigr\rangle \nonumber \\
&~~~~+
\textstyle
\contraction[1ex]{\bigl\langle{\frac{1}{2!}}i\int d^5x_1\bigl(-\frac{1}{4}\kappa\eta_{MN}}{h}
{\tilde{\partial}^{*N}\bar{\psi}i\Gamma^M\psi\bigr)i\int d^5x_2\bigl(\frac{1}{4}\kappa\eta_{M'N'}}{h}
\contraction[1.5ex]{\bigl\langle{\frac{1}{2!}}i\int d^5x_1\bigl(-\frac{1}{4}\kappa\eta_{MN}h\tilde{\partial}^{*N}}{\bar{\psi}}
{i\Gamma^M\psi\bigr)i\int d^5x_2\bigl(\frac{1}{4}\kappa\eta_{M'N'}h\bar{\psi}i\Gamma^{M'}\tilde{\partial}^{N'}}{\psi}
\contraction[2.6ex]{\bigl\langle{\frac{1}{2!}}i\int d^5x_1\bigl(-\frac{1}{4}\kappa\eta_{MN}h\tilde{\partial}^{*N}\bar{\psi}i\Gamma^M}{\psi}
{\bigr)i\int d^5x_2\bigl(\frac{1}{4}\kappa\eta_{M'N'}h}{\bar{\psi}}
\bigl\langle{\frac{1}{2!}}
i\int d^5x_1\bigl(-\frac{1}{4}\kappa\eta_{MN}h\tilde{\partial}^{*N}\bar{\psi}i\Gamma^M\psi\bigr)
i\int d^5x_2\bigl(\frac{1}{4}\kappa\eta_{M'N'}h\bar{\psi}i\Gamma^{M'}\tilde{\partial}^{N'}\psi\bigr)
\bigr\rangle \nonumber \\
&~~~~+
\textstyle
\contraction[1ex]{\bigl\langle{\frac{1}{2!}}i\int d^5x_1\bigl(-\frac{1}{4}\kappa\eta_{MN}}{h}
{\tilde{\partial}^{*N}\bar{\psi}i\Gamma^M\psi\bigr)i\int d^5x_2\bigl(-\frac{1}{4}\kappa\eta_{M'N'}}{h}
\contraction[1.5ex]{\bigl\langle{\frac{1}{2!}}i\int d^5x_1\bigl(-\frac{1}{4}\kappa\eta_{MN}h\tilde{\partial}^{*N}}{\bar{\psi}}
{i\Gamma^M\psi\bigr)i\int d^5x_2\bigl(-\frac{1}{4}\kappa\eta_{M'N'}h\tilde{\partial}^{*N'}\bar{\psi}i\Gamma^{M'}}{\psi}
\contraction[2.6ex]{\bigl\langle{\frac{1}{2!}}i\int d^5x_1\bigl(-\frac{1}{4}\kappa\eta_{MN}h\tilde{\partial}^{*N}\bar{\psi}i\Gamma^M}{\psi}
{\bigr)i\int d^5x_2\bigl(-\frac{1}{4}\kappa\eta_{M'N'}h\tilde{\partial}^{*N'}}{\bar{\psi}}
\bigl\langle{\frac{1}{2!}}
i\int d^5x_1\bigl(-\frac{1}{4}\kappa\eta_{MN}h\tilde{\partial}^{*N}\bar{\psi}i\Gamma^M\psi\bigr)
i\int d^5x_2\bigl(-\frac{1}{4}\kappa\eta_{M'N'}h\tilde{\partial}^{*N'}\bar{\psi}i\Gamma^{M'}\psi\bigr)
\bigr\rangle \nonumber \\
&=\frac{(i\kappa)^2}{2}\int d^5x_1\int d^5x_2 \nonumber \\
&~~~~
\times\biggl[
-\frac{1}{16}D_{MNM'N'}(\hat{x}_1,\hat{x}_2)
\mathrm{tr}\bigl\{\Gamma^MS^N(\hat{x}_1,\hat{x}_2)\Gamma^{M'}S^{N'}(\hat{x}_2,\hat{x}_1)\bigr\} \nonumber \\
&~~~~~~~~~~
-\frac{1}{16}D_{MNM'N'}(\hat{x}_1,\hat{x}_2)
\mathrm{tr}\bigl\{\Gamma^MS^{NN'}(\hat{x}_1,\hat{x}_2)\Gamma^{M'}S(\hat{x}_2,\hat{x}_1)\bigr\} \nonumber \\
&~~~~~~~~~~
+\frac{1}{16}\eta_{M'N'}\eta^{P'Q'}D_{MNP'Q'}(\hat{x}_1,\hat{x}_2)
\mathrm{tr}\bigl\{\Gamma^MS^N(\hat{x}_1,\hat{x}_2)\Gamma^{M'}S^{N'}(\hat{x}_2,\hat{x}_1)\bigr\} \nonumber \\
&~~~~~~~~~~
+\frac{1}{16}\eta_{M'N'}\eta^{P'Q'}D_{MNP'Q'}(\hat{x}_1,\hat{x}_2)
\mathrm{tr}\bigl\{\Gamma^MS^{NN'}(\hat{x}_1,\hat{x}_2)\Gamma^{M'}S(\hat{x}_2,\hat{x}_1)\bigr\} \nonumber \\
&~~~~~~~~~~
-\frac{1}{16}D_{MNM'N'}(\hat{x}_1,\hat{x}_2)
\mathrm{tr}\bigl\{\Gamma^MS(\hat{x}_1,\hat{x}_2)\Gamma^{M'}S^{N'N}(\hat{x}_2,\hat{x}_1)\bigr\} \nonumber \\
&~~~~~~~~~~
-\frac{1}{16}D_{MNM'N'}(\hat{x}_1,\hat{x}_2)
\mathrm{tr}\bigl\{\Gamma^MS^{N'}(\hat{x}_1,\hat{x}_2)\Gamma^{M'}S^N(\hat{x}_2,\hat{x}_1)\bigr\} \nonumber \\
&~~~~~~~~~~
+\frac{1}{16}\eta_{M'N'}\eta^{P'Q'}D_{MNP'Q'}(\hat{x}_1,\hat{x}_2)
\mathrm{tr}\bigl\{\Gamma^MS(\hat{x}_1,\hat{x}_2)\Gamma^{M'}S^{N'N}(\hat{x}_2,\hat{x}_1)\bigr\} \nonumber \\
&~~~~~~~~~~
+\frac{1}{16}\eta_{M'N'}\eta^{P'Q'}D_{MNP'Q'}(\hat{x}_1,\hat{x}_2)
\mathrm{tr}\bigl\{\Gamma^MS^{N'}(\hat{x}_1,\hat{x}_2)\Gamma^{M'}S^N(\hat{x}_2,\hat{x}_1)\bigr\} \nonumber \\
&~~~~~~~~~~
+\frac{1}{16}\eta_{MN}\eta^{PQ}D_{PQM'N'}(\hat{x}_1,\hat{x}_2)
\mathrm{tr}\bigl\{\Gamma^MS^N(\hat{x}_1,\hat{x}_2)\Gamma^{M'}S^{N'}(\hat{x}_2,\hat{x}_1)\bigr\} \nonumber \\
&~~~~~~~~~~
+\frac{1}{16}\eta_{MN}\eta^{PQ}D_{PQM'N'}(\hat{x}_1,\hat{x}_2)
\mathrm{tr}\bigl\{\Gamma^MS^{NN'}(\hat{x}_1,\hat{x}_2)\Gamma^{M'}S(\hat{x}_2,\hat{x}_1)\bigr\} \nonumber \\
&~~~~~~~~~~
-\frac{1}{16}\eta_{MN}\eta_{M'N'}\eta^{PQ}\eta^{P'Q'}D_{PQP'Q'}(\hat{x}_1,\hat{x}_2)
\mathrm{tr}\bigl\{\Gamma^MS^N(\hat{x}_1,\hat{x}_2)\Gamma^{M'}S^{N'}(\hat{x}_2,\hat{x}_1)\bigr\} \nonumber \\
&~~~~~~~~~~
-\frac{1}{16}\eta_{MN}\eta_{M'N'}\eta^{PQ}\eta^{P'Q'}D_{PQP'Q'}(\hat{x}_1,\hat{x}_2)
\mathrm{tr}\bigl\{\Gamma^MS^{NN'}(\hat{x}_1,\hat{x}_2)\Gamma^{M'}S(\hat{x}_2,\hat{x}_1)\bigr\} \nonumber \\
&~~~~~~~~~~
+\frac{1}{16}\eta_{MN}\eta^{PQ}D_{PQM'N'}(\hat{x}_1,\hat{x}_2)
\mathrm{tr}\bigl\{\Gamma^MS(\hat{x}_1,\hat{x}_2)\Gamma^{M'}S^{N'N}(\hat{x}_2,\hat{x}_1)\bigr\} \nonumber \\
&~~~~~~~~~~
+\frac{1}{16}\eta_{MN}\eta^{PQ}D_{PQM'N'}(\hat{x}_1,\hat{x}_2)
\mathrm{tr}\bigl\{\Gamma^MS^{N'}(\hat{x}_1,\hat{x}_2)\Gamma^{M'}S^N(\hat{x}_2,\hat{x}_1)\bigr\} \nonumber \\
&~~~~~~~~~~
-\frac{1}{16}\eta_{MN}\eta_{M'N'}\eta^{PQ}\eta^{P'Q'}D_{PQP'Q'}(\hat{x}_1,\hat{x}_2)
\mathrm{tr}\bigl\{\Gamma^MS(\hat{x}_1,\hat{x}_2)\Gamma^{M'}S^{N'N}(\hat{x}_2,\hat{x}_1)\bigr\} \nonumber \\
&~~~~~~~~~~
-\frac{1}{16}\eta_{MN}\eta_{M'N'}\eta^{PQ}\eta^{P'Q'}D_{PQP'Q'}(\hat{x}_1,\hat{x}_2)
\mathrm{tr}\bigl\{\Gamma^MS^{N'}(\hat{x}_1,\hat{x}_2)\Gamma^{M'}S^N(\hat{x}_2,\hat{x}_1)\bigr\}
\biggr] \nonumber \\
&=\frac{(i\kappa)^2}{2!}\int d^5x_1\int d^5x_2 \nonumber \\
&~~~~~
\times\frac{1}{16}\bigl\{-D_{MNM'N'}(\hat{x}_1,\hat{x}_2)+\eta_{M'N'}\eta^{P'Q'}D_{MNP'Q'}(\hat{x}_1,\hat{x}_2) \nonumber \\
&~~~~~~~~~~~~~~
+\eta_{MN}\eta^{PQ}D_{PQM'N'}(\hat{x}_1,\hat{x}_2)
-\eta_{MN}\eta_{M'N'}\eta^{PQ}\eta^{P'Q'}D_{PQP'Q'}(\hat{x}_1,\hat{x}_2)\bigr\} \nonumber \\
&~~~~~
\times\mathrm{tr}\bigl\{\Gamma^MS^N(\hat{x}_1,\hat{x}_2)\Gamma^{M'}S^{N'}(\hat{x}_2,\hat{x}_1)
+\Gamma^MS^{N'}(\hat{x}_1,\hat{x}_2)\Gamma^{M'}S^N(\hat{x}_2,\hat{x}_1) \nonumber \\
&~~~~~~~~~~~~~~
+\Gamma^MS^{NN'}(\hat{x}_1,\hat{x}_2)\Gamma^{M'}S(\hat{x}_2,\hat{x}_1)
+\Gamma^MS(\hat{x}_1,\hat{x}_2)\Gamma^{M'}S^{N'N}(\hat{x}_2,\hat{x}_1)\bigr\}.
\end{align}
Then, we can express $\hat{V}_{\mathrm{Fig}.\,\ref{fig:ge}}^{2\mathchar`-\mathrm{loop}}$ as
the first equality in the expression (\ref{5depge}), and it becomes
\begin{align}
\label{5depc2}
&-i\hat{V}_{\mathrm{Fig}.\,\ref{fig:ge}}^{2\mathchar`-\mathrm{loop}} \nonumber \\
&=\frac{(i\kappa)^2}{2!}\int \frac{d^5k}{(2\pi)^5}\int \frac{d^5p}{(2\pi)^5}\int \frac{d^5q}{(2\pi)^5}
(2\pi)^5\delta^{(5)}(\hat{k}+\hat{p}-\hat{q}) \nonumber \\
&~~~~~
\times\frac{1}{16}\bigl\{-\mathscr{D}_{MNM'N'}+\eta_{M'N'}\eta^{P'Q'}\mathscr{D}_{MNP'Q'} \nonumber \\
&~~~~~~~~~~~~~~
+\eta_{MN}\eta^{PQ}\mathscr{D}_{PQM'N'}
-\eta_{MN}\eta_{M'N'}\eta^{PQ}\eta^{P'Q'}\mathscr{D}_{PQP'Q'}\bigr\}
\mathrm{tr}\bigl\{\Gamma^M\Gamma^S\Gamma^{M'}\Gamma^{S'}\bigr\} \nonumber \\
&~~~~~
\times\frac{i}{\hat{k}^2+i\epsilon}
\biggl(
\frac{i\tilde{p}^N\tilde{p}_S}{\hat{\tilde{p}}^2+i\epsilon}\frac{i\tilde{q}^{N'}\tilde{q}_{S'}}{\hat{\tilde{q}}^2+i\epsilon}
+\frac{i\tilde{p}^{N'}\tilde{p}_S}{\hat{\tilde{p}}^2+i\epsilon}\frac{i\tilde{q}^N\tilde{q}_{S'}}{\hat{\tilde{q}}^2+i\epsilon}
+\frac{i\tilde{p}^N\tilde{p}^{N'}\tilde{p}_S}{\hat{\tilde{p}}^2+i\epsilon}\frac{i\tilde{q}_{S'}}{\hat{\tilde{q}}^2+i\epsilon}
+\frac{i\tilde{p}_S}{\hat{\tilde{p}}^2+i\epsilon}\frac{i\tilde{q}^{N'}\tilde{q}^N\tilde{q}_{S'}}{\hat{\tilde{q}}^2+i\epsilon}
\biggr) \nonumber \\
&=i\kappa^2\int \frac{d^5p_{\mathrm{E}}}{(2\pi)^5}\int \frac{d^5q_{\mathrm{E}}}{(2\pi)^5}
\biggl\{
\frac{11}{12}\frac{1}{(\hat{\tilde{p}}_{\mathrm{E}}-\hat{\tilde{q}}_{\mathrm{E}})^2}
+\frac{5}{3}\frac{\hat{\tilde{p}}_{\mathrm{E}}\hat{\tilde{q}}_{\mathrm{E}}}
{(\hat{\tilde{p}}_{\mathrm{E}}-\hat{\tilde{q}}_{\mathrm{E}})^2\hat{\tilde{p}}_{\mathrm{E}}^2}
+\frac{3}{4}\frac{(\hat{\tilde{p}}_{\mathrm{E}}\hat{\tilde{q}}_{\mathrm{E}})^2}
{(\hat{\tilde{p}}_{\mathrm{E}}-\hat{\tilde{q}}_{\mathrm{E}})^2\hat{\tilde{p}}_{\mathrm{E}}^2\hat{\tilde{q}}_{\mathrm{E}}^2}
\biggr\}.
\end{align}
The above expression (\ref{5depc2}) can be rewritten as the second equality in the expression (\ref{5depge}),
using the  change of variables $\hat{q}_{\mathrm{E}}=\hat{k}_{\mathrm{E}}+\hat{p}_{\mathrm{E}}$
($\Leftrightarrow\hat{\tilde{q}}_{\mathrm{E}}=\hat{k}_{\mathrm{E}}+\hat{\tilde{p}}_{\mathrm{E}}$)
and the identity:
\begin{align}
\int \frac{d^5p_{\mathrm{E}}}{(2\pi)^5}\int \frac{d^5q_{\mathrm{E}}}{(2\pi)^5}
\frac{(\hat{\tilde{p}}_{\mathrm{E}}\hat{\tilde{q}}_{\mathrm{E}})^2}
{(\hat{\tilde{p}}_{\mathrm{E}}-\hat{\tilde{q}}_{\mathrm{E}})^2\hat{\tilde{p}}_{\mathrm{E}}^2\hat{\tilde{q}}_{\mathrm{E}}^2}
=\int \frac{d^5p_{\mathrm{E}}}{(2\pi)^5}\int \frac{d^5q_{\mathrm{E}}}{(2\pi)^5}
\frac{1}{2}
\biggl\{
\frac{2\hat{\tilde{p}}_{\mathrm{E}}\hat{\tilde{q}}_{\mathrm{E}}}
{(\hat{\tilde{p}}_{\mathrm{E}}-\hat{\tilde{q}}_{\mathrm{E}})^2\hat{\tilde{p}}_{\mathrm{E}}^2}
-\frac{\hat{\tilde{p}}_{\mathrm{E}}\hat{\tilde{q}}_{\mathrm{E}}}
{\hat{\tilde{p}}_{\mathrm{E}}^2\hat{\tilde{q}}_{\mathrm{E}}^2}
\biggr\}.
\end{align}

\end{document}